\documentclass[journal]{IEEEtran}
\ifCLASSINFOpdf
\usepackage{graphicx}
\else
   \usepackage[dvips]{graphicx}
\fi
\usepackage{amsmath, amsfonts, url, booktabs}

\begin{document}

\title{Advanced Signal Analysis in Detecting Replay Attacks for Automatic Speaker Verification Systems}

\author{Lee Shih Kuang
\thanks{Lee Shih Kuang is an independent researcher from Taiwan,\\
(e-mail: lee@shihkuanglee.com).}}

\maketitle

\begin{abstract}
This study proposes novel signal analysis methods for replay speech detection in automatic speaker verification (ASV) systems.
The proposed methods---arbitrary analysis (AA), mel scale analysis (MA), and constant Q analysis (CQA)---are inspired by the calculation of the Fourier inversion formula.
These methods introduce new perspectives in signal analysis for replay speech detection by employing alternative sinusoidal sequence groups.
The efficacy of the proposed methods is examined on the ASVspoof 2019 \& 2021 PA databases with experiments, and confirmed by the performance of systems that incorporated the proposed methods; the successful integration of the proposed methods and a speech feature that calculates temporal autocorrelation of speech (TAC) from complex spectra strongly confirms it.
Moreover, the proposed CQA and MA methods show their superiority to the conventional methods on efficiency (approximately 2.36 times as fast compared to the conventional constant Q transform (CQT) method) and efficacy, respectively, in analyzing speech signals, making them promising to utilize in music and speech processing works.
\end{abstract}


\IEEEpeerreviewmaketitle

\section{Introduction}
The finite Fourier transform \cite{Wong2011} served as the cornerstone in signal processing to analyze the frequency composition of discrete-time signals.
While effective in many applications, including speech dereverberation (weighted prediction error) \cite{5547558} and speaker verification (mel-frequency cepstral coefficients) \cite{Kinnunen2004}, spectra with a linear scale may not be optimal (spectra calculated by constant Q transform (CQT) \cite{TODISCO2017516}) to capture the desired characteristics from signals.

\vfill

Previous research \cite{Lee2211:Detecting} has demonstrated that collecting autocorrelation data from both temporal (single spectral signal) and spatial (audio channels) domains in the same time effectively captures replay attacks, which present unique challenges compared to synthetic or converted speech detection in automatic speaker verification (ASV) systems.
Yet the proposed feature---temporal autocorrelation of speech (TAC)---is not compatible with nonlinear spectra that are calculated by established methods such as melspectrogram \cite{brian_mcfee-proc-scipy-2015}; TAC is calculated from complex spectra.

\vfill

Meanwhile, a series of challenges, named ASVspoof \cite{wu15e_interspeech, kinnunen17_interspeech, todisco19_interspeech, yamagishi21_asvspoof}, is established to promote the development of countermeasures to protect ASV systems from spoofing attacks.
ASVspoof first focused on attacks with synthetic and converted speech \cite{wu15e_interspeech}, then paid attention to replay attacks \cite{kinnunen17_interspeech}.
Since ASVspoof 2019 \cite{WANG2020101114}, the challenge addresses replay attacks as the physical access (PA) task, and it remained a separate task in ASVspoof 2021 \cite{10155166}.

\newpage

This study\footnote{This study is derived from my unpublished manuscript.}\cite{lee2024arbitrary} introduces novel signal analysis methods---arbitrary analysis (AA), mel scale analysis (MA), and constant Q analysis (CQA)---that are inspired by the calculation of the Fourier inversion formula; new insights on signal analysis that are relevant to replay speech detection are presented through alternative sinusoidal sequence groups.
The efficacy of the proposed methods is shown by:
\begin{enumerate}
  \item Successful integration with TAC
  \item Superior performance of systems incorporated them
  \item Desired characteristics captured by them
\end{enumerate}

\vfill

The results demonstrate that the proposed methods not only match but, in some cases, excel over conventional methods in terms of efficiency and effectiveness.
Specifically, the CQA method offers significant computational advantages over the traditional CQT method, while the MA method shows a superior ability in capturing human speech characteristics.

\vfill

The remainder of this paper is structured as follows:\newline
Section \ref{sec:2} describes the inspiration and complete details of the proposed methods.
Section \ref{sec:3} presents the experimental setup and evaluation methodology.
Section \ref{sec:4} discusses the results and their implications.
Sections \ref{sec:5} and \ref{sec:6} provide concluding remarks and directions for future work, respectively.

\section{Proposed Methods} \label{sec:2}

\subsection{Inspiration}
Calculations of the Fourier inversion formula \cite{Wong2011} inspired the proposed methods; here we denote the Fourier matrix in the finite Fourier transform $\mathcal{F}_{\mathbb{Z}_N}:L^2(\mathbb{Z}_N){\ }{\rightarrow}{\ }L^2(\mathbb{Z}_N)$ as $\Omega_N$
\begin{equation*}
\Omega_N =
\begin{pmatrix}
1      & 1              & 1                 & \dots  & 1 \\
1      & \omega_N       & \omega_N^2        & \dots  & \omega_N^{N-1} \\
1      & \omega_N^2     & \omega_N^4        & \dots  & \omega_N^{2(N-1)} \\
\vdots & \vdots         & \vdots            & \ddots & \vdots \\
1      & \omega_N^{N-1} & \omega_N^{2(N-1)} & \dots  & \omega_N^{(N-1)(N-1)}
\end{pmatrix}.
\end{equation*}
The finite Fourier transform $\hat{z}{\ }{\in}{\ }L^2(\mathbb{Z}_N)$ is calculated by
\begin{equation} \label{eq:1}
\hat{z} = \Omega_N z,
\end{equation}  
and the reconstruction of the signal $z{\ }{\in}{\ }L^2(\mathbb{Z}_N)$ is calculated with the Fourier inversion formula as follows
\begin{equation*}
z = \Omega_N^* \frac{1}{N} \Omega_N z.
\end{equation*}
Since frequency components are calculated to determine the characteristics (amplitude and phase) of sinusoidal sequences for reconstruction independently, it is inspiring to calculate the spectrum with other groups of sinusoidal sequences.

\newpage

What follows are the three proposed methods\footnote{\url{github.com/shihkuanglee/ADFA}} Vanilla, MA, and CQA; spectra used in this study are calculated with the proposed methods and the finite Fourier transform, as shown in Equation \ref{eq:1}.

\subsection{Vanilla}
Calculating the spectrum with sinusoidal sequences ranging from zero to Nyquist frequency linearly on angular frequencies is set as a vanilla method and named arbitrary analysis (AA); the sinusoidal sequences are
\begin{equation*}
\Omega_A =
\begin{pmatrix}
1          & 1        & 1          & \dots  & \omega_0^{N-1} \\
1          & \omega_1 & \omega_1^2 & \dots  & \omega_1^{N-1} \\
\omega_a^0 & \omega_a & \omega_a^2 & \dots  & \omega_a^{N-1} \\
\vdots     & \vdots   & \vdots     & \ddots & \vdots \\
1          & -1       & 1          & \dots  & \omega_{F-1}^{N-1}
\end{pmatrix},
\end{equation*}
where
\begin{equation*}
\omega_{a} = e^{-{\pi}ia/(F-1)},\quad a=0,1,\dots,F-1.
\end{equation*}
${F}$ is an arbitrary natural number to assign the number of components in the spectrum.

\subsection{Mel scale Analysis}
Mel scale analysis refers to calculating spectrum by using sinusoidal sequences with mel scale distances on angular frequencies from zero frequency to Nyquist frequency;\break the sinusoidal sequences are shown as follows
\begin{equation*}
\Omega_M =
\begin{pmatrix}
1          & 1        & 1          & \dots  & \omega_0^{N-1} \\
1          & \omega_1 & \omega_1^2 & \dots  & \omega_1^{N-1} \\
\omega_a^0 & \omega_a & \omega_a^2 & \dots  & \omega_a^{N-1} \\
\vdots     & \vdots   & \vdots     & \ddots & \vdots \\
1          & -1       & 1          & \dots  & \omega_{F-1}^{N-1}
\end{pmatrix},
\end{equation*}
where
\begin{equation*}
\omega_a = e^{-{\pi}i\frac{2}{f_s}Mel(\frac{f_s}{2}\frac{a}{F-1})},\quad a=0,1,\dots,F-1.
\end{equation*}
$Mel$ function converts values in hertz to the mel scale;\break$f_s$ stands for the sampling frequency of the signal $z$.

\subsection{Constant Q Analysis}
The constant Q analysis aims to calculate a spectrum with constant Q; the sinusoidal sequences are
\begin{equation*}
\Omega_Q =
\begin{pmatrix}
\omega_{F-1}^0 &\omega_{F-1} &\omega_{F-1}^2 &\dots  &\omega_{F-1}^{N-1}\\
\vdots         &\vdots       &\vdots         &\ddots &\vdots \\
\omega_a^0     &\omega_a     &\omega_a^2     &\dots  &\omega_a^{N-1}\\
1              &\omega_1     &\omega_1^2     &\dots  &\omega_1^{N-1}\\
1              &-1           &1              &\dots  &\omega_0^{N-1}
\end{pmatrix},
\end{equation*}
where
\begin{equation*}
\omega_a = e^{-i{\pi}/Q^a},\quad Q=\sqrt[B]{b},\quad a=F-1,\dots,1,0.
\end{equation*}
$Q$ stands for the constant Q that makes the relative positions of the pattern (such as musical sounds consisting of harmonic components) constant in the spectrum \cite{brown1991calculation}; $B$ represents the number of components per octave when base $b=2$.

\newpage
\vfill

\section{Experiments} \label{sec:3}
All systems\footnote{\url{github.com/shihkuanglee/RD-LCNN}} shown in this study are evaluated with standard metrics (lower is better) EER and min t-DCF metrics \cite{kinnunen18b_odyssey, kinnunen2020tandem} from ASVspoof challenges.
In order to verify the solidity of implementations, baseline systems \texttt{DFT} and \texttt{TAC} are built first, then realize the systems incorporating the proposed methods.
Equal numbers of trials are performed for systems with the same size; models are selected according to their performance (equal error rate (EER) as the primary metric) on \texttt{2019-dev} set, then evaluated on \texttt{2019-eval} and \texttt{2021-eval} sets.

\vfill

\subsection{Systems}

\subsubsection{Speech Features}
Log spectra and TAC \cite{Lee2211:Detecting} are adopted.\newline
Systems \texttt{Ceps} and \texttt{CQT} take cepstrogram and log spectrogram as speech features, respectively.

\subsubsection{Configurations}
Systems \texttt{CQT}, \texttt{DFT}, \texttt{AA}, \texttt{MA}, and \texttt{CQA} configured a Blackman window of length 1724 and frame shift 128 (dimensions [863, 600]); \texttt{Ceps}, \texttt{TAC}, \texttt{ATAC}, \texttt{MTAC} and \texttt{QTAC} configured a Blackman window of length 1024 and frame shift 256 (dimensions [513, 600] and [513, 16]); \texttt{CQT}, \texttt{CQA}, and \texttt{QTAC} have similar frequencies (around 15.625 Hz) on the lowest components.

\subsubsection{Model}
The light convolutional neural network (LCNN) architecture is chosen to use in this study since it showed robustness against spoofing attacks on three challenges \cite{lavrentyeva17_interspeech, lavrentyeva19_interspeech, tomilov21_asvspoof}, and was the most representative architecture (rank 2 in both tracks) in ASVspoof 2019 \cite{WANG2020101114}.
The identical architecture was used in the system \texttt{T01} (rank 4) \cite{10155166} from the ASVspoof 2021 PA task.

\vfill

\begin{table}
\scriptsize
\centering
\caption{Computation time of bona fide trials in 2019-dev set}
\label{tab:time}
\begin{tabular}{cl}
\toprule
\textbf{Method} & \textbf{Time (Seconds)} \\
\midrule
CQT & 266 \\
CQA & \textbf{113} ($-57.5$\%) \\
\bottomrule
\end{tabular}
\end{table}

\subsection{Results}

\subsubsection{Table I}
Experimental results highlight the superior efficiency of the proposed CQA method compared to conventional CQT\footnote{\url{github.com/asvspoof-challenge/2021/}}; the experiments are done on the same device, and the programs are single-threaded and written in the same language.
The CQA method calculates spectra with a constant Q approximately 2.36 times as fast compared to the conventional CQT method.

\subsubsection{Table II}
The effectiveness of the proposed methods in replay speech detection is strongly confirmed by the matched performance of systems (\texttt{CQT} and \texttt{CQA}), and the progressive improvement of systems \texttt{TAC}, \texttt{ATAC}, \texttt{MTAC}, and \texttt{QTAC} on \texttt{2019-dev} and \texttt{2019-eval}; it is founded on successful integrations and superior performance.
Furthermore, system \texttt{MA} demonstrates its capability to the unseen condition \texttt{2021-eval} set, excelling \texttt{Ceps} (top single system on both \texttt{2019-dev} and \texttt{2019-eval} to the best of my knowledge), suggesting its potential in general replay speech detection.

\newpage

\begin{table}
\scriptsize
\centering
\caption{Performance and size of systems}
\label{tab:systems}
\begin{tabular}{r@{\hspace{10pt}}r@{\hspace{8pt}}l@{\hspace{6pt}}l@{\hspace{8pt}}l@{\hspace{6pt}}l@{\hspace{8pt}}l@{\hspace{6pt}}l}
\toprule
\multicolumn{2}{c}{} & \multicolumn{2}{c}{\textbf{2019-dev}} & \multicolumn{2}{c}{\textbf{2019-eval}} & \multicolumn{2}{c}{\textbf{2021-eval}} \\
\cmidrule(r){3-8}
\multicolumn{1}{c}{\textbf{System}} & \multicolumn{1}{l}{\textbf{Size}} & \textbf{t-DCF} & \textbf{EER} & \textbf{t-DCF} & \textbf{EER} & \textbf{t-DCF} & \textbf{EER} \\
\midrule
\cite{Lee2211:Detecting}               TAC & 1.29M & 0.0863 & 3.152 & 0.1560 & 5.882 & $\approx$ 1 & $\approx$ 50 \\
\cite{Lee2211:Detecting, lee2022study} CQT & 40.8M & 0.0096 & 0.374 & 0.0130 & 0.514 & 0.9761 & 41.21 \\
\cite{Lee2211:Detecting, lee2022study} Ceps & 24.88M &\textbf{0.0039} & \textbf{0.129} & \textbf{0.0105} & \textbf{0.370} & \textbf{0.9288} & \textbf{36.75} \\
\cmidrule(r){2-8}
 DFT & 40.8M & \textbf{0.0031} & \textbf{0.111} &         0.0192  &         0.653  & 0.9729 & 42.07 \\
  AA & 40.8M &         0.0034  &         0.168  &         0.0127  &         0.481  & 0.9769 & 40.07 \\
  MA & 40.8M &         0.0040  &         0.148  &         0.0188  &         0.631  & \textbf{0.8548} & \textbf{35.50} \\
 CQA & 40.8M &         0.0056  &         0.222  & \textbf{0.0127} & \textbf{0.442} & 0.9989 & 44.48 \\
\cmidrule(r){2-8}
 TAC & 1.29M &         0.0548  &         1.963  &         0.0975  &         3.626  & 0.9055 & 42.58 \\
ATAC & 1.29M &         0.0478  &         1.630  &         0.0955  &         3.362  & \textbf{0.9032} & 38.19 \\
MTAC & 1.29M &         0.0447  &         1.704  &         0.0776  &         2.931  & 0.9738 & \textbf{37.13} \\
QTAC & 1.29M & \textbf{0.0414} & \textbf{1.442} & \textbf{0.0714} & \textbf{2.619} & 0.9479 & 38.97 \\
\bottomrule
\end{tabular}
\end{table}

\section{Discussions} \label{sec:4}

\subsection{Finite Fourier Transform versus Vanilla}
Credibility is considered first when the experiment begins; it is offered by the evaluation of the systems of finite Fourier transform (\texttt{DFT}, \texttt{TAC}) and Vanilla (\texttt{AA}, \texttt{ATAC}) in Table \ref{tab:systems}.
Mathematically, calculating the spectra as speech features is identical for both finite Fourier transform and Vanilla systems due to the even number frame length used in the analysis; however, additional signal processing techniques are applied to the systems \texttt{DFT} (\textbf{Spectrogram}\footnote{\url{pytorch.org/audio/main/transforms.html}}) and \texttt{TAC} (\textbf{stft}\footnote{\url{github.com/fgnt/nara_wpe/blob/master/nara_wpe/utils.py}}) before calculating the spectra, resulting in distinct performance.
The training strategy may contribute the most to Vanilla systems in outperforming finite Fourier transform systems; stochastic gradient descent and balanced sampling of (spoofed / original) trials are applied along with minimum signal processing to the trials during model training to maximize the capabilities of the LCNN architecture; the strategy is also applied to \texttt{T01} \cite{10155166}.

\subsection{CQT and CQA}
\begin{figure}[!h]
  \centering
  \includegraphics[width=0.48\textwidth]{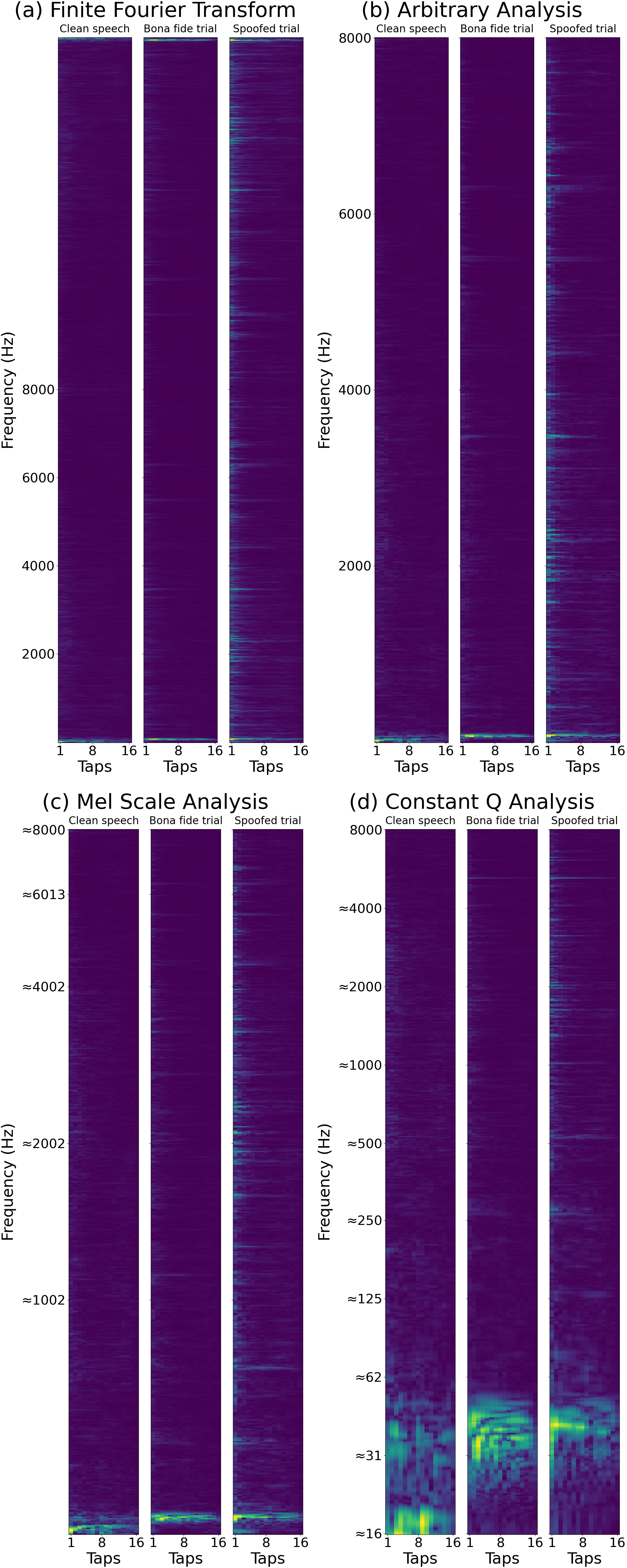}
  \caption{TACs (plotted in log1p-scale) comparing analytical methods for clean speech (p262\_227 from \cite{VCTK}), bona fide trial (PA\_D\_0004063, p262\_227 with simulated reverberation) and spoofed trial (PA\_D\_0024255, PA\_D\_0004063 with replay attack).}
  \label{fig:TACs}
\end{figure}

\begin{figure*}
  \centering
  \includegraphics[width=\textwidth]{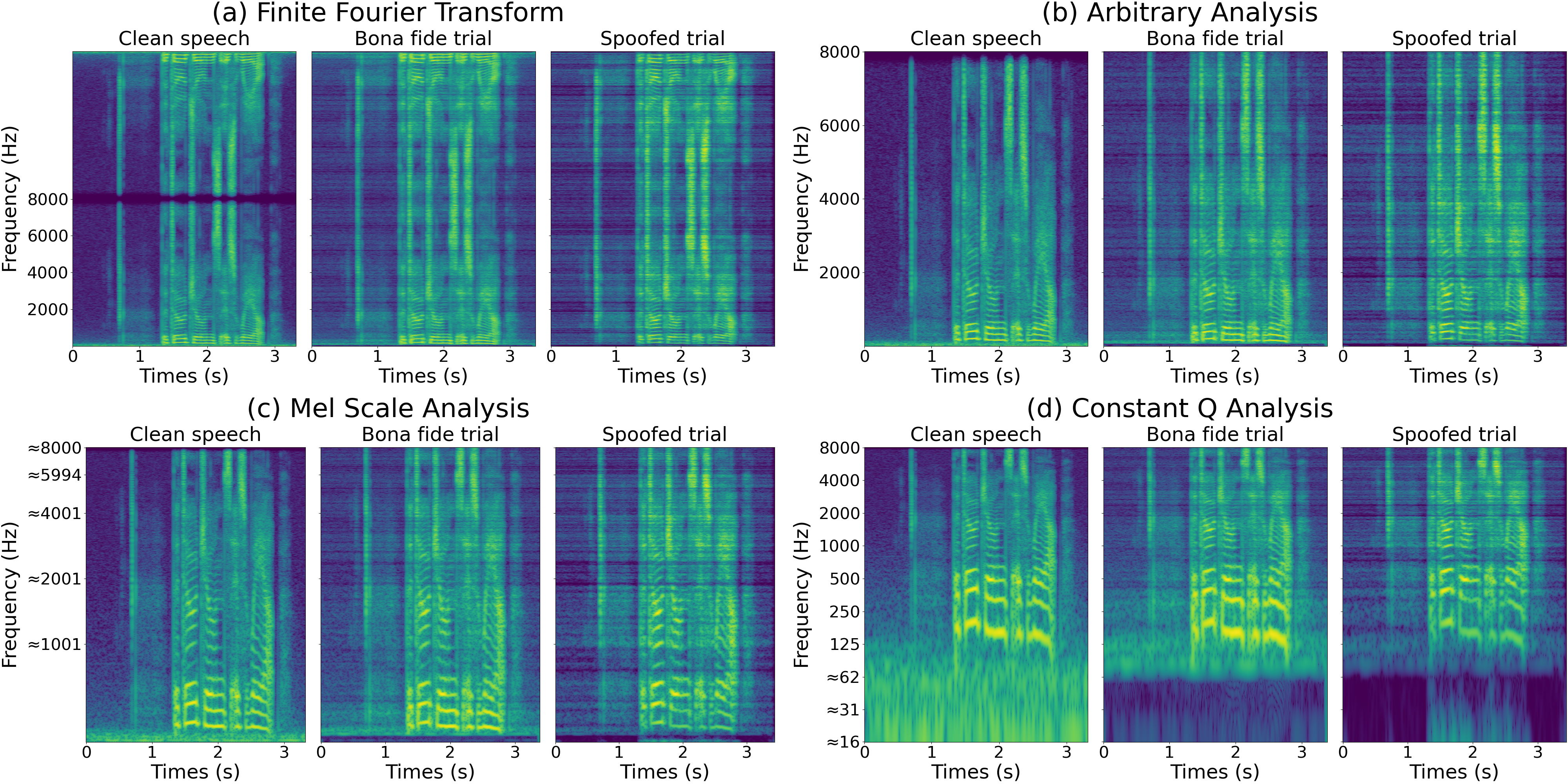}
  \caption{Speech features (plotted in log-scale) comparing analytical methods for clean speech (p262\_227 from \cite{VCTK}), bona fide trial (PA\_D\_0004063, p262\_227 with simulated reverberation) and spoofed trial (PA\_D\_0024255, PA\_D\_0004063 with replay attack). Note how the spectra of clean speech is smeared by simulated reverberation and replay attack, and how their trajectories become apparent in panel (d).}
  \label{fig:ADFAs}
\end{figure*}

Tables \ref{tab:time} \& \ref{tab:systems} demonstrate the efficacy of the proposed methods through computation time and systems' performance.
Since both TAC and the constant Q spectrum are calculated independently in the frequency domain, the effectiveness of the CQA method in analyzing speech signals is confirmed by the performance of the \texttt{QTAC} system.
Moreover, the proposed CQA method is significantly faster than the conventional CQT method \cite{TODISCO2017516} in achieving the constant Q spectra, as shown in Table \ref{tab:time}, making it feasible to compute training data online. 
In addition to the effectiveness and efficiency of the CQA method in detecting replay attacks, it is easier for humans to separate spoofed trails from genuine trails with the help of the proposed method; Figures \ref{fig:TACs} and \ref{fig:ADFAs} present the visualization of speech features as progress in magnifying the trajectories of replay attack, making them more distinguishable for us.
However, unlike the performance of systems \texttt{TAC}, \texttt{ATAC}, \texttt{MTAC}, and \texttt{QTAC} on \texttt{2019-dev} and \texttt{2019-eval} sets, the progressive improvement on \texttt{2021-eval} set stopped at the systems \texttt{MA} and \texttt{MTAC}, suggesting that the optimal nonlinear analysis for general replay speech detection may lie in an alternative beyond the constant Q used in this study.

\vfill
\newpage

\newpage

\subsection{MA}
\begin{figure}[h]
  \centering
  \includegraphics[width=0.48\textwidth]{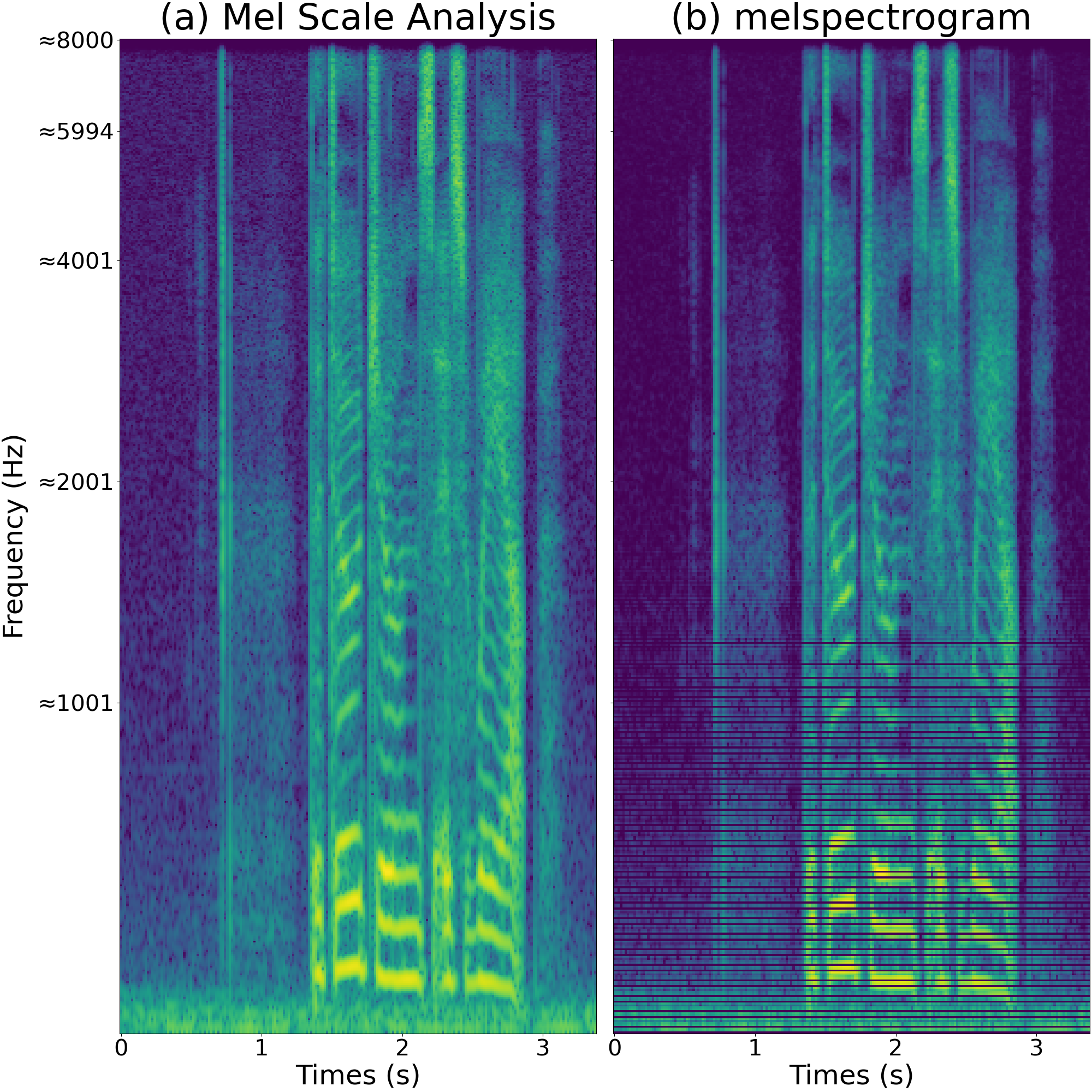}
  \caption{The clean speech calculated with (a) MA and (b) melspectrogram.}
  \label{fig:MAs}
\end{figure}

Clear visions into human speech characteristics are offered by the proposed MA method, as shown by Figure \ref{fig:MAs} (a) \& (b).\break
The traditional method\footnote{\url{librosa.org/doc/latest/generated/librosa.feature.melspectrogram.html}} generates the mel scale spectrum with empty spectral components and pixelation on the spectrum when calculating with a larger number of components; it not only degrades the quality of the mel spectrum as a speech feature but also limits its capability in integrating with other techniques such as TAC for replay speech detection.
The MA method, by contrast, demonstrates its potential in capturing human speech characteristics for general replay speech detection, as evidenced by the performance of the \texttt{MA} system on the \texttt{2021-eval} set (as shown in Table \ref{tab:systems}).
This method calculates frequency components directly using the sinusoidal sequence group with the mel scale, resulting in a complete spectrum that is compatible with TAC.

\section{Conclusions} \label{sec:5}
New methods for signal analysis are proposed in this study for ASV systems; the proposed methods AA, MA, and CQA are carefully examined with spoofing attacks of the physical access scenarios from the ASVspoof 2019 / ASVspoof 2021.\break
Their efficacy is presented by visualizations of speech features and experimental results as shown in Figures and Tables; the integration of them and the TAC feature strongly confirms it.
Moreover, the capability of the MA method is uncovered by visual comparison with the traditional method and featured in experimental results with the leading performance on general replay speech detection for ASV systems, resulting in a fruitful method for capturing human speech characteristics.

\section{Future Work} \label{sec:6}
Revisiting studies that involved the use of finite Fourier transform with the proposed methods is planned; such as text-to-speech synthesis with MA and music processing with CQA.

\bibliographystyle{ieeetr}
\bibliography{refs.bib}

\end{document}